# End to End Simulation of AO-assisted coronagraphic differential imaging: estimation of performance for SPHERE

Anthony Boccaletti[a], Marcel Carbillet[b], Thierry Fusco[c], David Mouillet[d],
Maud Langlois[e], Claire Moutou[e], Kjetil Dohlen[e] [*]

[a] LESIA, Observatoire de Paris, 5 pl. J. Janssen, F-92195 Meudon, France;
[b] FIZEAU, Université de Nice-Sophia Antipolis, Parc Valrose, F-06108 Nice, France
[c] ONERA-DOTA, BP 72, 29 avenue de la Division Leclerc, F-92322 Châtillon Cedex, France
[d] LAOG, BP 53, 414 Rue de la Piscine, F-38041 Grenoble Cedex, France
[e] LAM, Observatoire de Marseille-Provence, 38 rue F. Joliot-Curie, F-13388 Marseille, France

## ABSTRACT

SPHERE (Spectro Polarimetric High contrast Exoplanet REsearch), the planet finder instrument for the VLT is designed to study relatively bright extrasolar giant planets around young or nearby stars. SPHERE is a set of three instruments fed by the same AO-system, two of them share the same coronagraph. This complex system has been modeled with Fourier Optics to investigate the performance of the whole instrument. In turns, this end-to-end model was useful to analyze the sensitivity to various parameters (WFE, alignment of the coronagraph, differential aberrations) and to put some specifications on the sub-systems. This paper presents some example of sensitivity analysis and some contrast performance of the instruments as a function of the flux for the main observing mode of SPHERE: the Dual Band Imaging (DBI), equivalent to the Spectral Differential Imaging technique.

**Keywords:** Coronagraphy, High Contrast Imaging, Exoplanets,

## 1. INTRODUCTION

Since the discovery of the first exoplanets by mean of high precision Radial Velocity measurements (Mayor & Queloz 1995, Marcy & Butler 1996), a lot of efforts have been put on direct detection techniques with the goal to study this stage of the stellar formation. Unambiguous detections of low mass objects within the planetary regime were obtained with current imaging instruments (Chauvin et al. 2005, Neuhaüser et al. 2005). However, a more systematic study is requiring dedicated instruments for high contrast imaging. The first series of such advanced instruments will see first light beginning of 2011 both at the VLT (Beuzit et al. 2006) and Gemini (Macintosh et al. 2006). The study reported hereafter was carried out in the phase B of SPHERE (Spectro Polarimetric High contrast Exoplanet Research) the European planet finder instrument for the VLT.

SPHERE is composed of an extreme AO system (Fusco et al. 2006) serving an IR channel containing two differential imagers, one in the form of a dual beam camera (IRDIS, Dohlen et al. 2006) and one as an Integral Field Spectrograph (IFS, Claudi et al. 2006). The Visible channel where the wavefront sensing is performed is sharing light with a differential polarimeter (ZIMPOL, Schmid et al. 2005).

The principle of differential imaging was formulated by Racine et al. (1999) for two images taken simultaneously in the goal to reduce the speckle noise (atmospheric and static). The Spectral Differential Imaging (SDI) technique intensively studied by Marois et al. (2000) is limited by phase chromatism (the two images are formed at two different wavelengths) and differential aberrations (the beam pass through different optics). SDI was later implemented at CFHT (Marois et al. 2003) and at VLT (Lenzen et al. 2005).

A general approach of differential imaging with a coronagraph is presented in Cavarroc et al. 2006 and fundamental limitations are derived. It is demonstrated that static aberrations upstream and downstream the coronagraph have different behavior and therefore must be considered separately in the modeling of the instrument. We will make use of this model in the following.

---

[*] Further author information: (Send correspondence to A.B.)
A.B.: E-mail: anthony.boccaletti@obspm.fr, Telephone: +33 (0)1 45 07 77 21

This paper will focus on one particular observing mode of SPHERE, the Dual Band Imaging (DBI), which is similar to SDI with a coronagraph and a set of couple of filters in the near IR.

## 2. PRINCIPLE OF THE SIMULATION

A first study of the SPHERE/DBI mode performance and sensitivity was reported by Boccaletti et al. (2006) at the time of the phase A. This paper presents further investigations based on a very detailed error budget. Simulations presented hereafter have been obtained with the simulation tool written with IDL under the CAOS "problem solving" environment (Carbillet et al. 2001, http://fizeau.unice.fr/caos/). The SPHERE simulation is implemented as a Software package (Carbillet et al. 2008, this conference) and is modeling the different parts of the instrument (IRDIS, IFS and ZIMPOL). Some results for these other sub-systems are reported in this conference (Thalmann et al. 2008, Claudi et al. 2008, Vigan et al. 2008).

We are considering a specific observing strategy for the DBI mode. As said above, the performance of SDI is limited by phase chromatism and differential aberrations. Such defects can be calibrated if one introduces some sort of "diversity" in the signal. To avoid the use of reference star, which is known to be very constraining, it is proposed to obtain a second observation of the same target star at a different position in the sky. DBI being operated in a pupil-stabilized mode, an off-axis object like a planet rotates in the field as the star moves across the sky. The contrast can be improved if the speckle pattern remains unchanged while the star is tracked. SPHERE being installed on the Nasmyth platform, the only moving parts are the telescope pupil in front of the instrument (but it is filtered at low frequencies by the AO system) and the ADC (downstream the AO beamsplitter), which may produce differential aberrations and chromaticity. However, if symmetrical observations with respect to meridian (same zenithal angle) are obtained, the ADC does not produce differential effects and the speckle pattern can be further rejected. This observing strategy is referred as the Double Difference (DD). Sensitivity and contrast plot will be assessed with this method even if additional gain can be brought with the Angular Differential Imaging (ADI, Marois et al. 2006).

The whole error budget we are introducing in the simulation is listed below :

*Atmospheric parameters*

- seeing = 0.85"
- wavefront outerscale = 25m
- turbulent layers :
    altitude [m] = 0, 1000, 10000
    $Cn^2$ [%] = 20, 60, 20
    wind speed (m/s) = 12.5, 12.5, 12.5
    direction [°] = 0, 45, 90
- mV = 8
- zenith angle = 30°
- instrumental jitter = 3mas

*AO parameters*

- imaging wavelength = 1.6 μm
- linear nbr of sub-apertures = 40
- integration time [s] = 0.000833
- loop delay [s] = 0.001
- loop gain = 0.466
- WFS wavelength [nm] = 650
- RON [e-] = 0.5
- dark current [e-/s] = 2
- spatial filtering = 0.52

*Static aberrations*

- UT4 M1, M2, M3 figures
- instrument = 34.5nm
- AO calibration = 7.4nm
- Fresnel propagration = 4.7nm
- beam shift = 8nm

*Temporal aberrations*

- defocus = 4nm
- pupil shear = 0.002 D
- pupil rotation = 0.1°
- differential rotation = 0nm
- differential beam shift = 0nm
- offset pointing = 0.5mas

The temporal aberrations refer to the defects in between the two observations of the star symmetrically from the meridian. Typical defects in a coronagraphic system arise at the level of the focal mask (defocus, offset pointing) and at the Lyot stop (shear or rotation). The parameters of temporal aberrations are corresponding to the specifications we have put on the system on the basis of our first analysis (Boccaletti et al. 2006). DBI simulations are obtained with the 4 Quadrant Phase Mask coronagraph (4QPM, Rouan et al. 2000) and the H2H3 pair of filter ($\lambda_2$=1.60 μm, $\lambda_1$=1.65μm).

We also account for ADC chromatic residuals that we model as an offset pointing on the coronagraph (typically a fraction of mas) and for the mis-alignment onto the coronagraph in between the 2 observations (0.5mas).

## 3. SENSITIVITY AT HIGH FLUX OF THE DBI MODE

In this section, we analyze the sensitivity of the DBI mode at high flux neglecting photon noise, background noise, read out noise and Flat Field, in order to estimate only the instrumental contrast (often call speckle noise or speckle limitation). In the following figures, we plot the averaged PSF radial profile and then the 5-σ detection level of the coronagraphic image (red), the Single Difference (SD, orange) and the Double Difference (DD, green).

### 3.1 Averaging of AO residuals

The AO system simulation delivers uncorrelated phase screens. We compared in Fig. 1 the contrast level of 100 iterations and 1000 iterations and found that they are nearly identical even at the Double Difference stage. The first case was therefore considered as a baseline for the simulations hereafter.

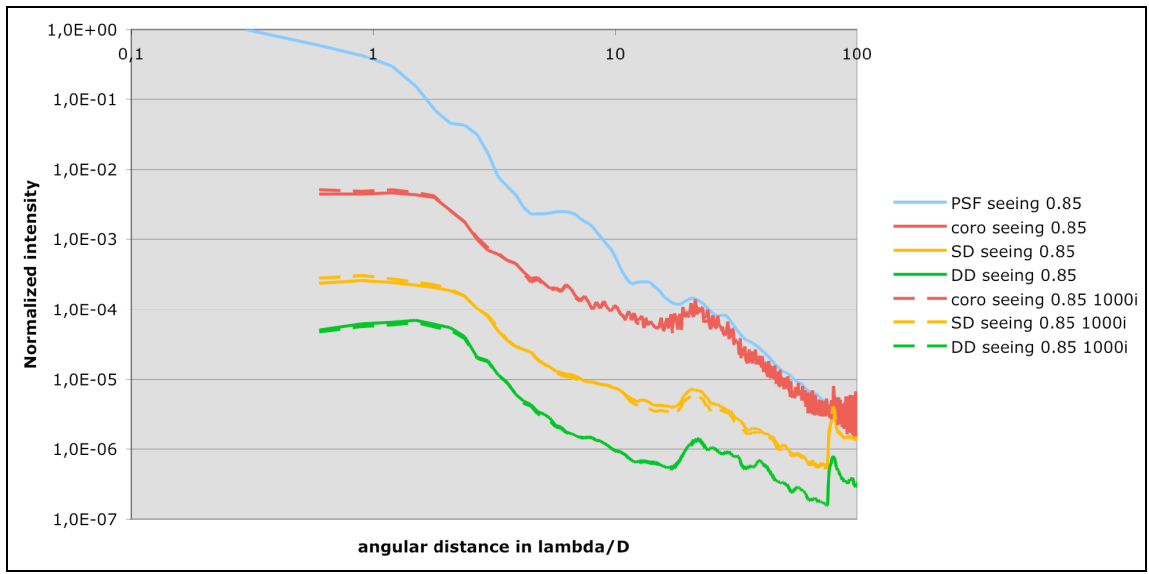

Figure 1: radial profiles comparing 100 and 1000 iterations.

High contrast imaging has to deal with 2 components (at first order), a speckled halo which is averaging over time and a static speckle pattern originating from static aberrations. Obviously, the static aberrations are not perfectly static but since they have a much longer lifetime than atmospheric residuals and also because we are performing simultaneous differential imaging we consider them as perfectly static. Since the speckle halo for infinite exposure time has zero variance and therefore should not contribute to the final signal to noise ratio, we consider that simulation can be made without the atmosphere as explained in Cavarroc et al. 2006. This assumption is valid for high flux only. If one has to consider photon noise, then the atmospheric residual has to be taken into account to correctly estimate the intensity level of the coronagraphic image. In this section, simulations are performed without the AO halo term.

### 3.2 Wavelength rescaling

Differential imaging at 2 wavelengths requires the images to be rescaled spatially in order to match the 2 speckle patterns and allow subtraction. Presently, no optimization of the rescaling has been implemented in the simulation and is therefore achieved with standard IDL routines. To evaluate the effect of the numerical errors we simulate a case where the 2 images have the same scale in order to avoid the rescaling. As expected, avoiding rescaling gives better contrast

(less than a factor of 2) but only in the inner region for r<5λ/D on the SD and r<3λ/D for the SD. Therefore, for the purpose of this paper, we did not attempt to optimize the spatial rescaling.

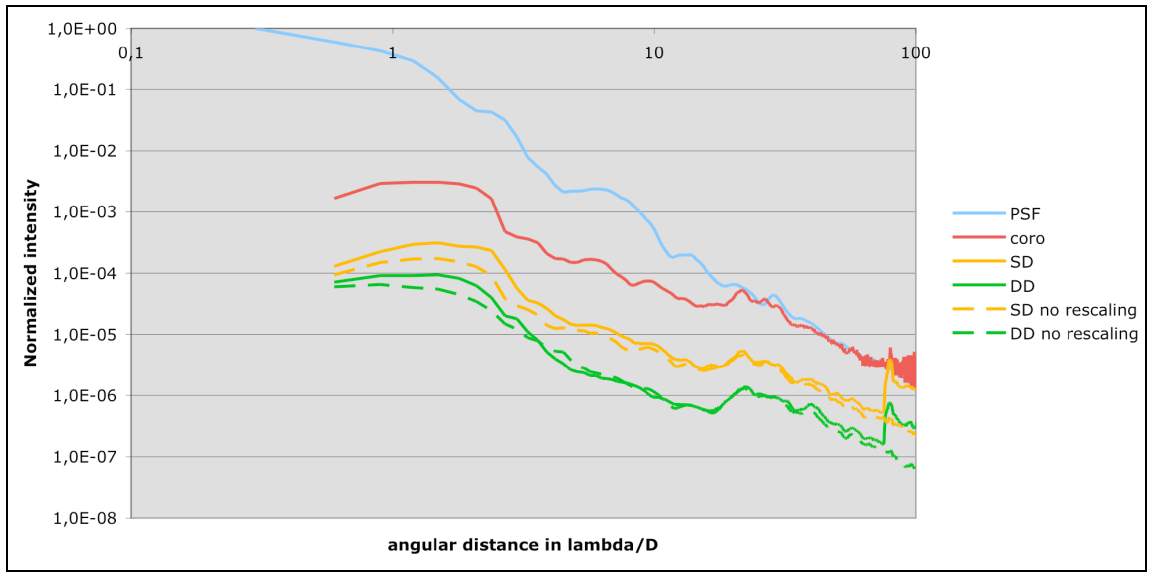

Figure 2: Radial profiles comparing the standard case with and without spatial rescaling

### 3.4 Static aberrations upstream and downstream coronagraph

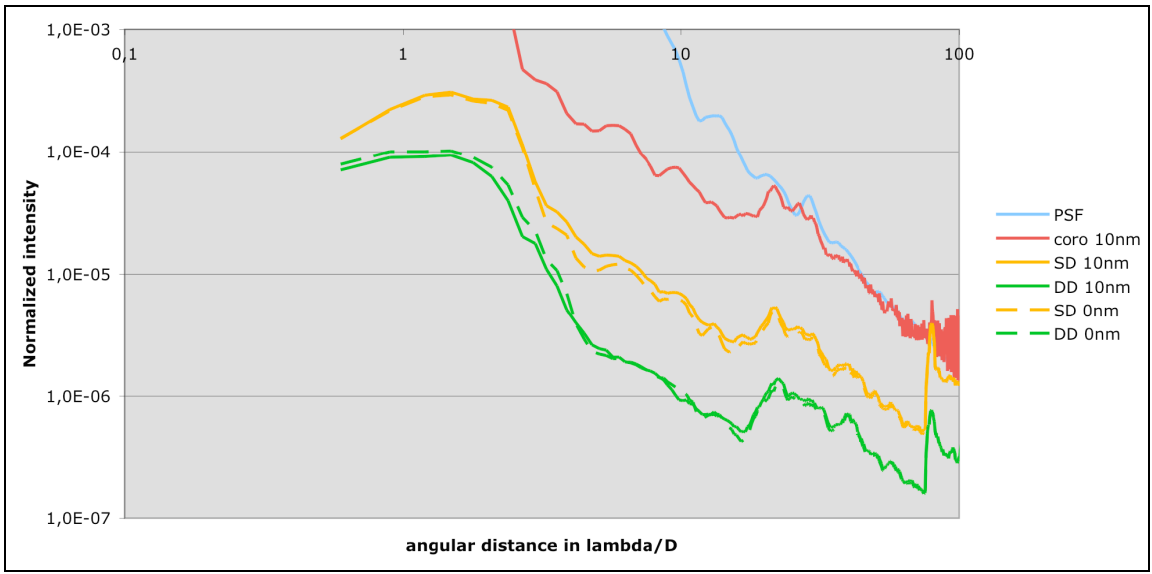

Figure 3: Radial profiles with 0nm and 10nm rms of differential aberrations downstream the coronagraph (36nm rms upstream).

As explained in Cavarroc et al. (2006), the final contrast of an SDI instrument is resulting from the combination of common aberrations upstream the coronagraph and differential aberrations dowstream. The standard assumptions for SPHERE are corresponding to 36nm and 10nm rms for respectively common and differential aberrations. Although differential aberrations were considered as critical, Fig. 3 clearly shows no significant improvement in between the current situation (10nm rms) and an ideal one (0nm rms) for the Double as well as the Single Difference. This effect is a direct consequence of chromatism due to the spectral separation in between filter. If filters could be made closer, we would have observed a progression of the SD contrast as differential aberrations decrease.

The Double Difference is able to reject most of the differential speckles and the chromatism indicating that differential aberrations are not a limitation for DBI. However, there is a net improvement when common aberrations are reduced by a factor of 2 (Fig. 4) which again agrees with the study of Cavarroc et al. (2006).

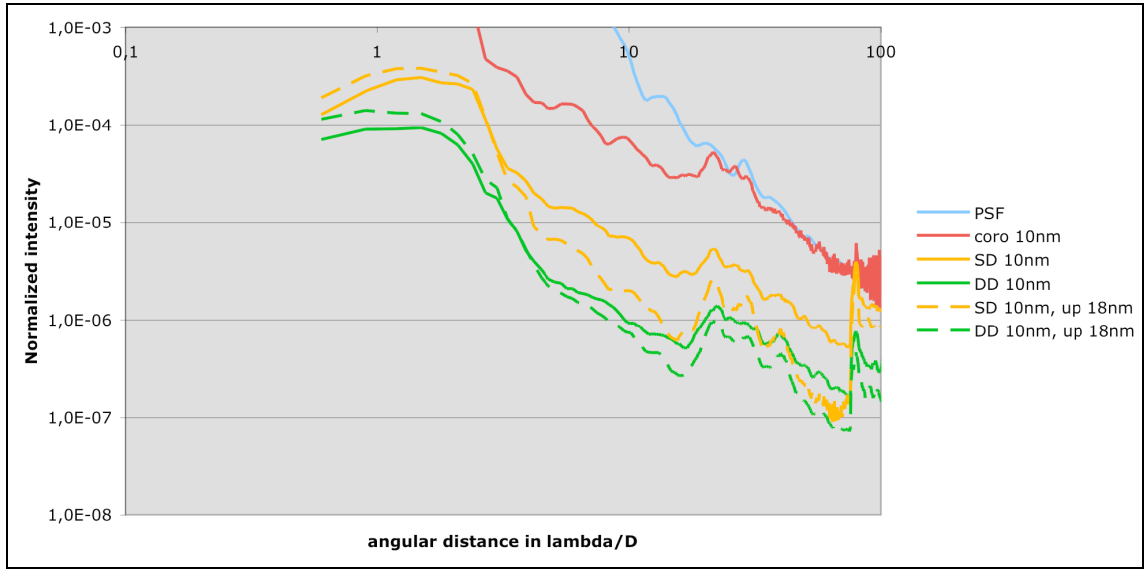

Figure 4: Radial profiles with 36nm and 18nm rms of static (common) aberrations upstream the coronagraph (10nm rms downstream).

## 3.5 Differential rotation

To be efficient, the Double Difference requires the two observations of the star to be obtained at the same zenithal angle but with a different field orientation. If the observations were obtained not symmetrically from the meridian, the ADC and the differential beam shift would introduce an aberration amoutning to 18nm rms. This would produce a complete decorrelation of the speckles resulting in a DD contrast equivalent to the SD profile as seen in Fig. 5.

## 3.6 Field rotation

The DBI mode is here operated in pupil stabilized mode which means that the field is rotating while most of the residual speckles are stable. Processing of the data requires a derotation of the field of view to sum-up the individual planet images. Two types of speckles are therefore left in the image: those which are perfectly static and those which are evolving while FOV is rotating. For the first ones the derotation is resulting in averaging the same speckle realization while the evolving speckles are also averaged but with several realizations. A simple simulation equivalent to ADI (Marois et al. 2006) was performed to quantify this effect. We considered short exposures on which the field rotation is frozen. Data processing (SD and DD) is performed on each of these individual exposures and then derotation is applied. We consider several rotation angle from 0° (the standard case presented above) to 180° with a sampling of 0.5° (which is needed to provide independent speckle realizations at the edge of the simulated field of view). The result presented in Fig. 6 shows that the contrast is naturally improved with the amplitude of the field rotation but has a radial dependence. This can be understood since for a given field rotation more independent speckle realizations are averaged at the edge of the field than near the centre of the field. Also, a field rotation of 25° is needed to observe a significant gain at short angular separations (2-3 $\lambda/D$). In the mid-frequencies region (10-20 $\lambda/D$) a gain of almost 10 in contrast can be expected for a large rotation of 180°.

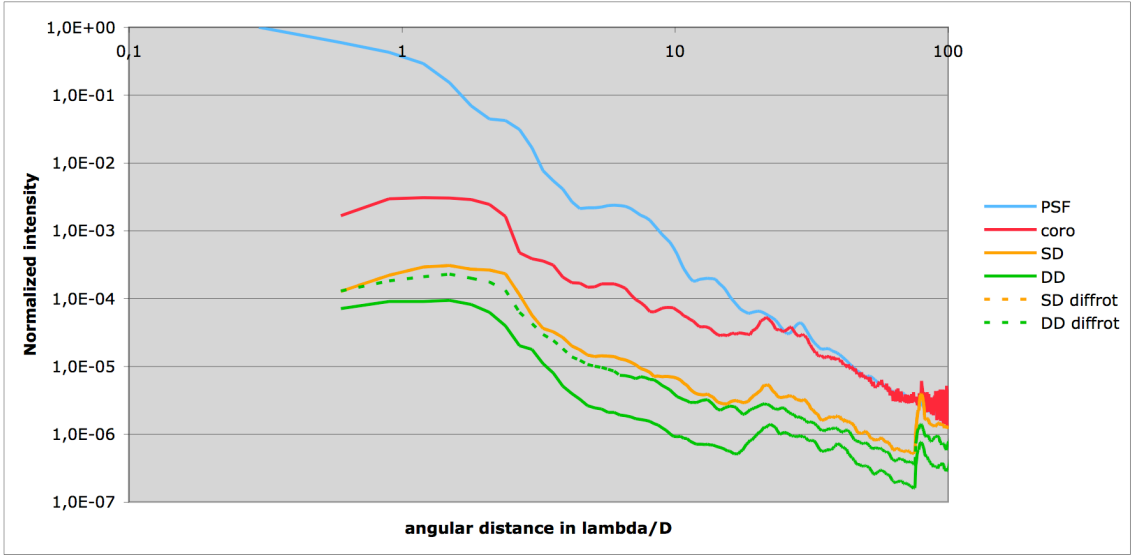

Figure 5: Radial profiles comparing the case of a reference star at the same parallactic angle with a reference at a zenithal angle of 15° from the star (differential rotating aberrations accounted).

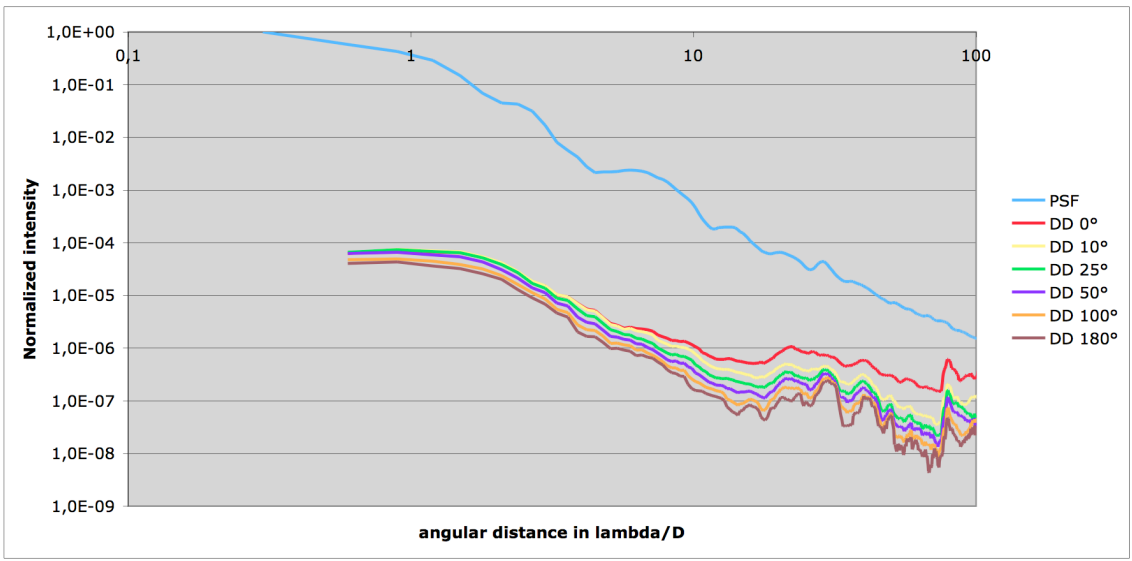

Figure 6: Radial profiles of the DD for several field rotations.

## 4. HIGH FLUX CONTRASTS

This section compares the achievable contrast for several DBI pairs of filters. IRDIS includes 10 pairs of filters but we only report for the main ones being Y1Y2 (0.97µm, 1.02µm), J2J3 (1.19µm, 1.27µm), H2H3 (1.60µm, 1.65µm), K1K2 (2.10µm, 2.25µm) in Fig. 7. Performance improves with wavelength except for the pair K1K2 which has a larger spectral separation than other DBI filters and therefore features more chromatism.

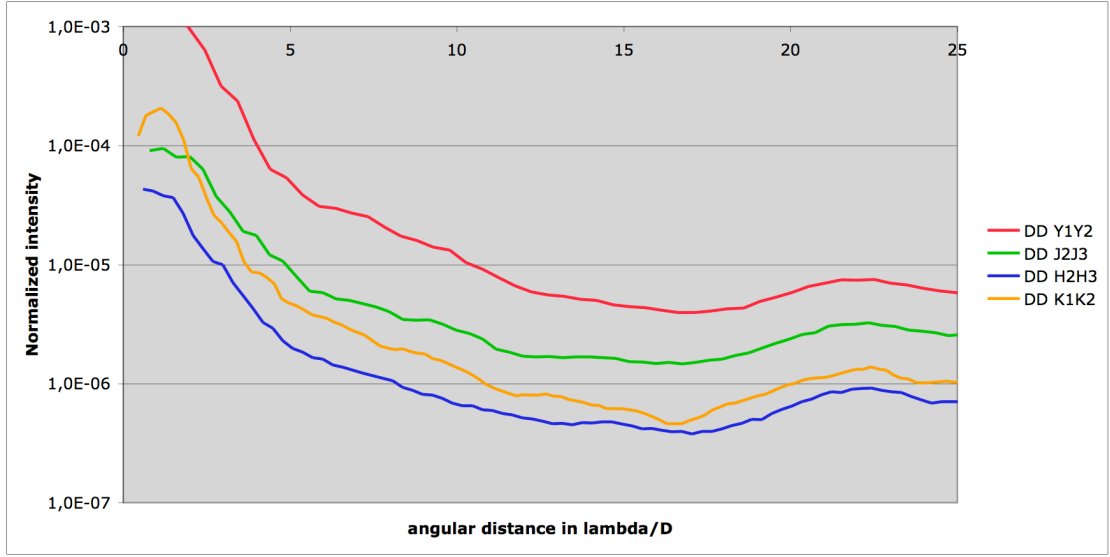

Figure 7: comparison of DD radial profiles for the different filters displayed linearly between 0 and 25 λ/D.

## 5. LOW FLUX CONTRASTS

### 5.1 Contrasts vs. exposure times

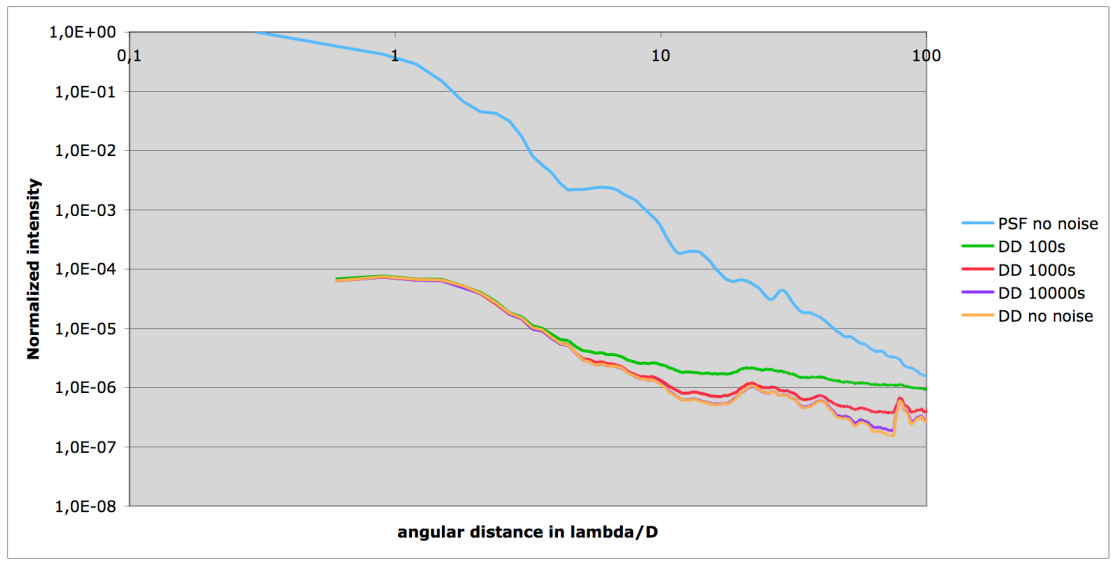

Figure 8: Radial profile for 100, 1000 and 10000s assuming a G0V at 10pc

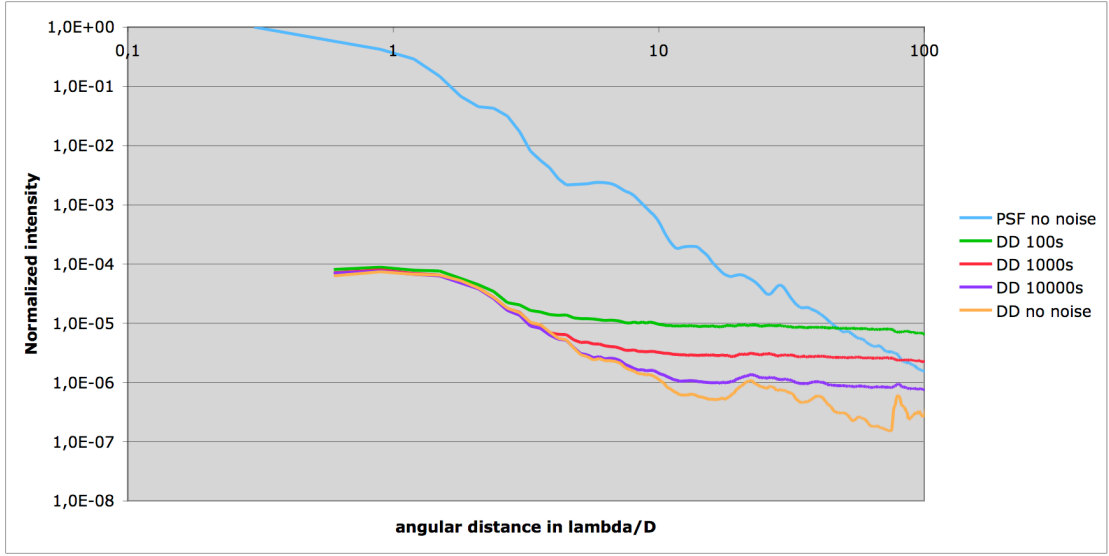

Figure 9: Radial profile for 100, 1000 and 10000s assuming a M0V at 10pc

Previous sections have presented the capabilities of SPHERE in DBI for high flux (or infinitely long exposure). The DD contrast profile therefore represents the ultimate performance. For an actual star, the intensity of the coronagraphic image sets the photon noise that propagates in the DBI process. The achievable contrast critically depends on the photon flux and hence on the integration time. In this section, we have plotted for several types of stars the DD profile as a function of the integration time. At this point, the simulation includes the AO halo (for a correct estimate of the intensity in the coronagraphic image), the photon noise, the background noise, the read out noise and Flat Field defects ($10^{-3}$).

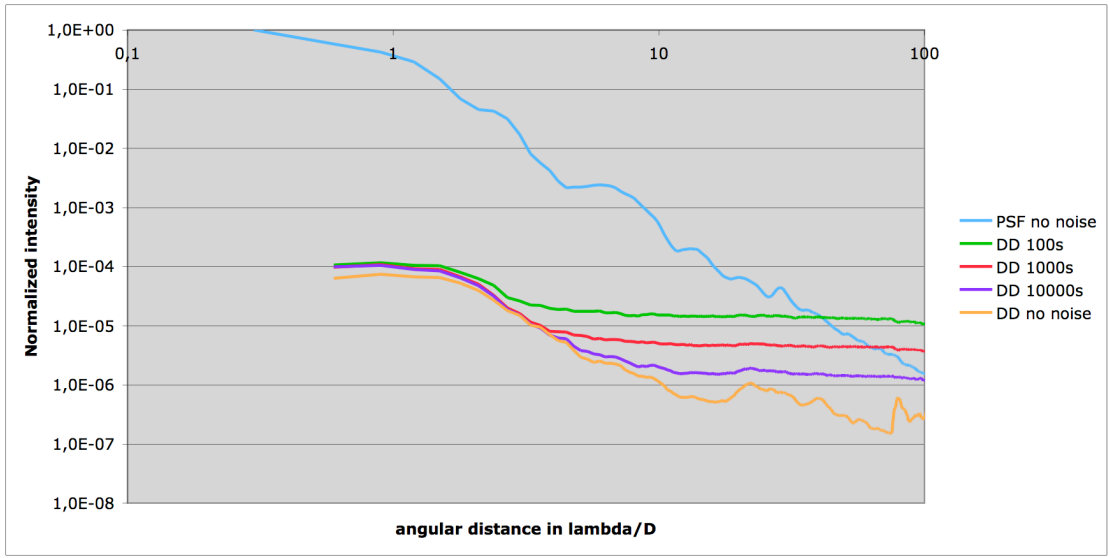

Figure 10: Radial profile for 100, 1000 and 10000s assuming a G0V at 40pc

Although the instrumental contras is achieved for a G0V star at 10pc in a few hundreds of seconds it is no longer the case for a fainter star (M0V) or at larger distances. The worst case presented here for a M0V at 40pc does not allow to reach a contrast higher than $10^5$ in 10000s. The photon noise limitation is therefore very critical and this level directly depends on the speckle halo behind the coronagraph. Assuming coronagraph behaves correctly, the limitation is therefore directly set by the capability of the AO system. SPHERE and GPI are designed to produce Strehl ratios of about 90% at H, which means that 10% of the light still remains in the coronagraphic image and in any case prevents to reach the instrumental contrast at low flux. It is very important for a project like SPHERE to estimate properly the performance at low flux since it is precisely the regime for which the star/planet ratio is the more favorable (either for later type stars or in young stellar associations located at distances larger than 40pc).

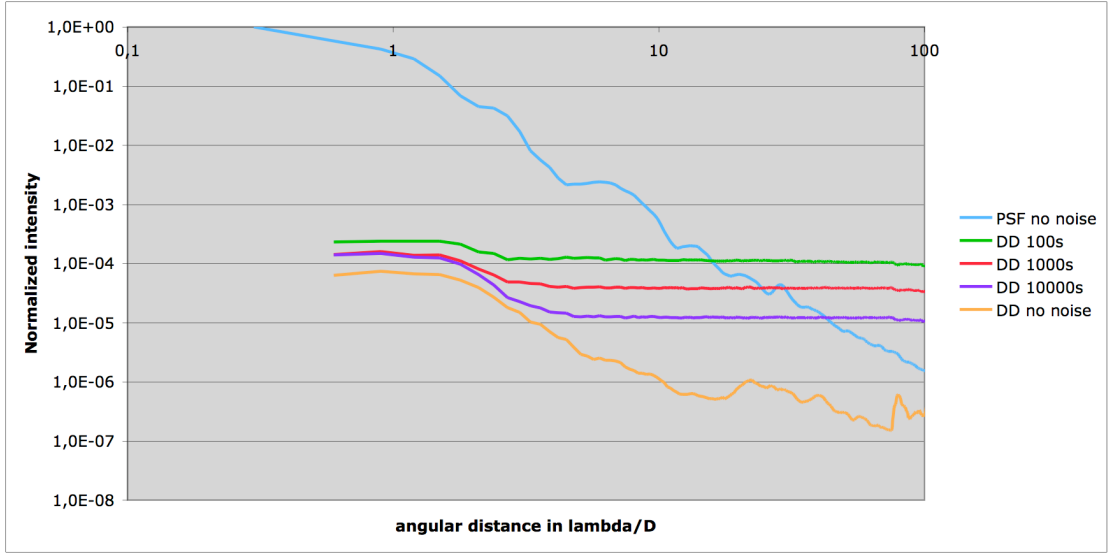

Figure 11: Radial profile for 100, 1000 and 10000s assuming a M0V at 40pc

## 5.2 Signal to noise ratio

In addition to the simulation of the on-axis star from which we derived the instrumental contrast in Sections 3 and 4 and the limitation from photon noise in Section 5.1, we also included 5 off-axis objects at various angular separations: 0.1", 0.2", 0.5", 1" and 2", in order to investigate the signal to noise ratio for different type of giant planets. The planets we are considering are listed in the next table together with corresponding masses as a function of the age.

Table 1: list of planets and correspondence between temperature, mass, and age.

| Planet temperature [K] / model | Corresponding Mass [MJ] / Age [Myr or Gyr] |
| --- | --- |
| 400 COND | 0.5MJ / 10Myr or 2MJ / 100Myr or 7MJ / 1Gyr or 15MJ / 5Gyr |
| 500 COND | 1MJ / 10My or 3MJ / 100My or 11MJ / 1Gy or 25MJ / 5Gy |
| 600 COND | 1.5MJ / 10My or 5MJ / 100My or 13MJ / 1Gy or 30MJ / 5Gy |
| 700 SETTLED | 2MJ / 10M or 6MJ / 100My or 18MJ / 1Gy or 35MJ / 5 Gy |
| 700 COND | 2MJ / 10My or 6MJ / 100My or 18MJ / 1Gy or 35MJ / 5 Gy |
| 900 SETTLED | 3MJ / 10My or 10MJ / 100My or 25MJ / 1Gy or 50MJ / 5Gy |
| 1000 COND | 4MJ / 10My or 11MJ / 100My or 30MJ / 1 Gy or 55MJ / 5Gy |
| 1200 SETTLED | 5MJ / 10My or 11MJ / 100My or 35MJ / 1Gy or 60MJ / 5Gy |

The intensity of each planet was determined and over-plotted on the detection limit. Figure 12 shows the level of detection for a G0V and M0V stars at 10pc observed in DBI with the filter pairs H2H3 and J2J3 for an integration time of 1 hour. As a result, a 400K planet is detectable at separation larger than 0.3-0.5" around an M0V while the limit is rather 500-600K for a G0V star. Other situations not presented here have been explored like for instance the possibility to take advantage of the reflected light from a planet angularly close to a bright and nearby star. In such a case we have obtained a detection limit of about 1 radius of Jupiter at separation of 0.2" independently of mass and temperature.

Therefore the limit of 1 mass of Jupiter could be approached with SPHERE in some very favorable conditions. Based on these results we are now able to carry out a better definition of the scientific program of SPHERE.

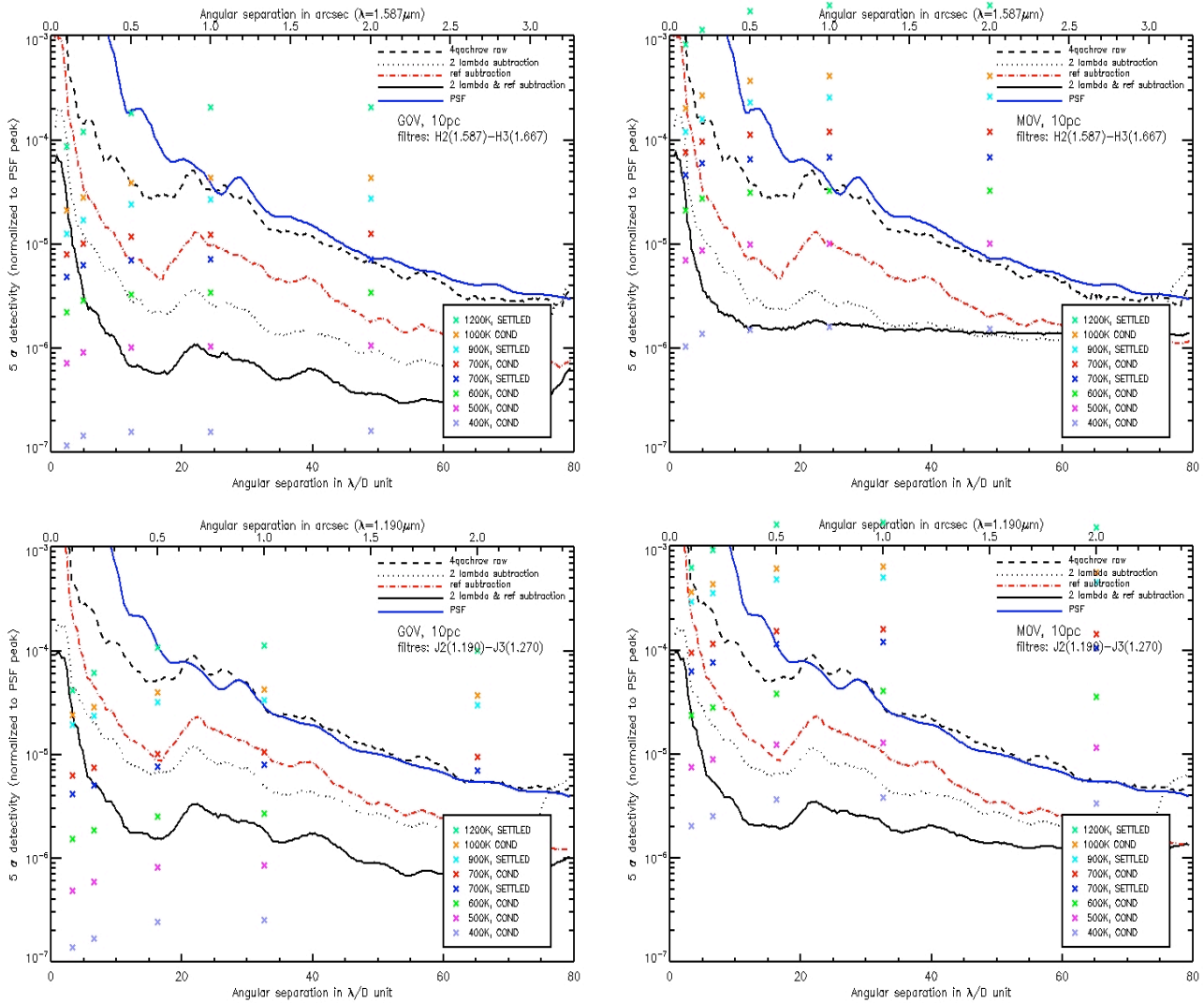

Figure 12: Estimated detectivity in H2H3 (top) and J2J3 (bottom) for an integration time of 1 hour assuming G0V (left) and M0V (right) stars at 10pc. The lower dotted and solid lines stand for the Single and Double Difference (Section 2).

## 6. CONCLUSION

The study presented in this paper was carried out in the framework of the SPHERE phase B. A previous analysis of sensitivity to system parameters and an estimate of contrast performance and signal to noise ratio were presented in an earlier paper (Boccaletti et al. 2006). Now, we have refined the AO simulation and the system parameters (budget for static as well as temporal aberrations) and we have estimate instrumental contrast, integration time and signal to noise ratio for planetary models. The simulation tool we have developed and which is detailed in a companion paper (Carbillet et al., this conference) has been proved very efficient to produce homogenous performance estimates of all the modes of SPHERE (dual band imaging, dual polarization imaging, integral field spectroscopy). We are now starting a phase of simulation to produce realistic data set in order to test data reduction algorithms (Smith et al. 2008, Mugnier et al. 2008).